\documentclass[10pt,letterpaper,copyedit]{article}
\usepackage{osajnl2}
\usepackage{epsf}
\usepackage{graphicx,epsf,subfigure}
\usepackage{amsmath,amsfonts,mathrsfs}

\begin{document}
\title{Digital in-line holography with an elliptical, astigmatic Gaussian beam : wide-angle reconstruction}

\author{N. Verrier, S. Coëtmellec, M. Brunel and D. Lebrun}
\address{ Groupe d'Optique et d'Optoélectronique, UMR-6614
CORIA, Av. de l'Université, \\76801 Saint-Etienne du Rouvray cedex,
France }

\author{A.J.E.M Janssen }
\address{  Philips Research Laboratories-Building WO-02,
Prof. Holstlaan 4,  5656 AE Eindhoven,  The Netherlands}

\email{coetmellec@coria.fr,\;a.j.e.m.janssen@philips.com}

\begin{abstract}
We demonstrate in this paper that the effect of object shift in an
elliptical, astigmatic Gaussian beam does not affect the optimal
fractional orders used to reconstruct the holographic image of a particle
or another opaque object in the field. Simulations and experimental
results are presented.
\end{abstract}


\section{Introduction}

Digital in-line holography (DIH) is widely used in microscopy
for biological applications \cite{Xu,Dubois}, in 3D Holographic Particle Image
Velocimetry (HPIV) for fluid mechanics studies\cite{Skarman} and in
refractometry \cite{Sebesta}. In most theoretical DIH studies, the
optical systems or the objects used, for example particles, are
considered to be centered on the optical axis \cite{Schnars,Shen}.
Nevertheless, in many practical applications, the systems and
objects are not necessarily centered. This is not a problem under
plane wave illumination. But more and more studies involve a diverging
beam, or astigmatic beams directly obtained at the output of fibered
laser diodes or astigmatic laser diodes. The position of the object
in the beam becomes particularly relevant. To reconstruct the image
of an object, a reconstruction parameter must be determined in order to find the best focus plane. For wavelet transformation, the
parameter is the scale factor \cite{Onural}. For Fresnel
transformation, the parameter is the distance $z$ from the object to
the quadratic sensor plane \cite{Onural_Scott} and for the fractional Fourier transformation, the parameters are the
fractional orders \cite{Coëtmellec,Fogret_Pellat-Finet}. It is
considered here that the same parameter can be used for all
positions of the object. Recently, an analytical solution of scalar
diffraction of an elliptical and astigmatic Gaussian beam (EAGB) by
a centered opaque disk under Fresnel approximation has been
proposed. By using the fractional Fourier transformation, a good
particle image reconstruction is obtained \cite{Nicolas}. However,
there are no recent publications in DIH theory demonstrating that reconstruction parameter is constant for all transverse
positions of the object in the field of the beam. This is, however,
particularly important if we want to develop wide-angle metrologies.

In this publication, the aim is to demonstrate that the same
fractional orders can be used to reconstruct an image of the
particle whatever its position in the field of the beam. We exhibit
the effect of the Gaussian beam on the reconstructed image. In the
first part of this publication, the model of the analytical solution
of scalar diffraction of an EAGB by a centered opaque disk is
revisited to take into account a decentered object. In the second
part, the definition of the fractional Fourier transformation is
recalled and this transformation is used to reconstruct the image of
the particle. It is in this part that we demonstrate that the same
orders can be used. Finally, we propose to illustrate our results by
simulations and experimental results.

\section{In-line Holography with an elliptic and astigmatic Gaussian beam}

\begin{figure}[t]
\centering
\includegraphics*[width=15cm]{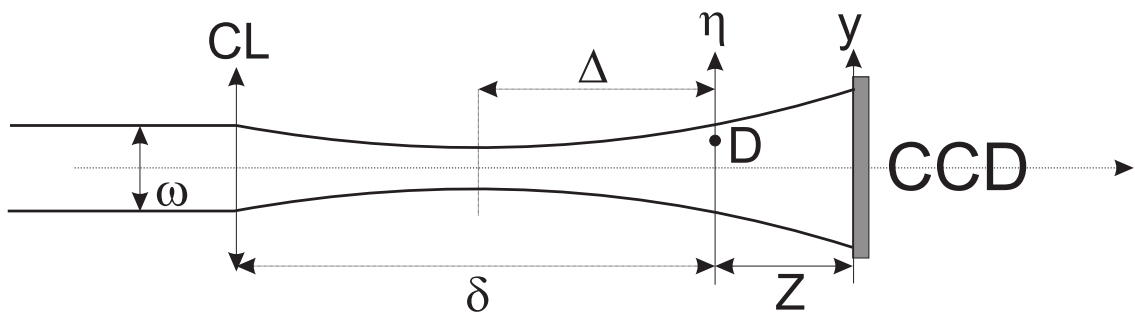}
\caption{Numerical and experimental optical set-up.}\label{fig1}
\end{figure}

The basic idea in DIH is to record by a CCD camera the intensity
distribution of the diffraction pattern of an object illuminated by
a continuous or pulsed wave \cite{Nicolas_2}. Figure (\ref{fig1})
represents the numerical and experimental set-up where all
parameters are identified. The incident Gaussian beam of diameter
$\omega$ crosses a plano-convex cylindrical lens. This
cylindrical lens acts only in the $\eta$-axis. Its focal length is
equal to $f_{\eta}=200mm$. After propagation in free space, the
elliptical and astigmatic Gaussian beam illuminates an opaque
particle. In the plane of the object, the beam widths along the
$\xi$-axis and $\eta$-axis are defined by $\omega_{\xi}$ and
$\omega_{\eta}$. The wavefront curvatures of the astigmatic beam are
denoted by $R_{q}$ with $q=\xi, \eta$. The CCD camera is located at
a distance $z$ from the opaque particle.

The basic model to describe the intensity distribution of the
diffracted beam by an object recorded by the CCD camera is the
integral of Kirchhoff-Fresnel given by the scalar integral for the
complex amplitude $A$ in the quadratic sensor plane:
\begin{equation}\label{holo:equ1}
A=\frac{\exp(i\frac{2\pi}{\lambda}z)}{i\lambda z}\int_{\mathbb{R}^2}
E_{T}(\xi,\eta)\,\exp\left(\frac{i\pi}{\lambda
z}\left[(\xi-x)^2+(\eta-y)^2\right]\right)\,d\xi d\eta,
\end{equation}
in which $E_{T}(\xi,\eta)$ is the product of the optical incident
beam, denoted $E(\xi,\eta)$, by the spatial transmittance of the
shifted opaque 2D-object, denoted $1-T(\xi-\xi_{0},\eta-\eta_{0})$.
The quadratic sensor records the intensity defined by $|A|^{2}$. If
we consider that the function $E_{T}$ is the product of an elliptic
and astigmatic Gaussian beam by an opaque disk of diameter $D$, then
\begin{equation}\label{holo:equ2}
E_{T}(\xi,\eta)=\underbrace{\exp\left[c_{\xi}
\xi^{2}+c_{\eta}\eta^{2}\right]}_{=E(\xi,\eta)}\cdot
\underbrace{[1-T(\xi-\xi_{0},\eta-\eta_{0})]}_{shifted\,\,object}.
\end{equation}
The complex coefficients $c_{\xi}$ and $c_{\eta}$ are
\begin{equation}\label{holo:equ3}
c_{\xi}=-\frac{1}{\omega_{\xi}^{2}}-i\frac{\pi}{\lambda R_{\xi}},
\quad c_{\eta}=-\frac{1}{\omega_{\eta}^{2}}-i\frac{\pi}{\lambda
R_{\eta}},
\end{equation}
For an opaque disk centered at the origin O, the transmittance
function $T(\xi,\eta)$ in the object plane is:
\begin{equation}\label{fonction_circ1}
T(\xi,\eta)=
  \begin{cases}
    1, & \text{$0<\sqrt{\xi^{2}+\eta^{2}}$ $<$ $D/2$}, \\
    1/2, & \text{$0<\sqrt{\xi^{2}+\eta^{2}}$ $=$ $D/2$}, \\
    0, & \text{$\sqrt{\xi^{2}+\eta^{2}}$ $>$ $D/2>0$}.
  \end{cases}
\end{equation}
>From Eqs. (\ref{holo:equ1}) and (\ref{holo:equ2}), the expression
for $A(x,y)$ can split into two integral terms, denoted $A_1$ and
$A_2$ so that:
\begin{equation}\label{holo:equ4}
A(x,y)=\frac{\exp(i\frac{2\pi}{\lambda}z)}{i}\left[A_{1}-A_{2}\right].
\end{equation}
The expression of these two terms is given in the Appendix (\ref{appA}).
The complex amplitude $A_1$ represents the propagation of the
incident beam without diffraction by the particle and $A_2$ contains
the diffraction by a pinhole of diameter $D$. After some development
detailed in the Appendix (\ref{appA}), the expression of $A_2$
becomes:
\begin{equation}\label{fonction_A2}
A_{2}=\frac{\pi D^2}{\lambda z}\exp\left[\Phi
(\xi_{0},\eta_{0})\right]\cdot \exp\left(
ic_z\left[(x-\xi_0)^2+(y-\eta_0)^2\right]\right) \cdot
\sum_{k=0}^{\infty}(-i)^{k}\varepsilon_{k}\,T_{k}(r,\gamma)\cos(2k\theta),
\end{equation}
with
\begin{equation}\label{fonction_Tk}
T_{k}(r,\gamma)=\sum_{p=0}^{\infty}\beta_{2k+2p}^{2k}(\delta)V_{2k+2p}^{2k}(r,\gamma),
\end{equation}
and
\begin{equation}\label{Vnm}
V_{2k+2p}^{2k}(r,\gamma)=\exp(i\gamma/2)\sum_{m=0}^{\infty}(2m+1)i^{m}
j_{m}\small{(\gamma/2)}\cdot
\sum_{l=\max(0,m-2k-p,p-m)}^{m+p}(-1)^{l}\omega_{ml}\frac{J_{2k+2l+1}(r)}{r}.
\end{equation}
The coefficients $\omega_{ml}$ are given explicitly in
\cite{Nicolas} and \cite{Janssen04}. Finally, the expression of
$A_1$ contains only the characteristics of the incident beam and
$A_2$ contains a shifted linear chirp function linked to the object
shift in the EAGB. The previous function is modulated by a series of
Bessel functions which constitutes the envelope of the amplitude
distribution of $A_2$.

A similar expression has been established in \cite{Nicolas} for a
particle located on the axis of the beam, \textit{i.e.} $\xi_0=0$
and $\eta_0=0$. This last expression is more general and can be
applied to a particle located everywhere in the field of the beam.

In our previous study, we demonstrated that a
fractional-order Fourier transformation allows a
particle located on the axis of the beam to be reconstructed. Unfortunately, the
expression of the diffraction pattern (and particularly $A_2$) is so
complex that it cannot be proved theoretically that a digital (or
optical) reconstruction leads effectively to an opaque disk
function. The demonstration is empirical. In addition, nothing
proves that the fractional orders used for the reconstruction would
be appropriate if the object were shifted transversely.

In the next developments, we will thus prove theoretically that the
reconstruction leads effectively to the object. We will then
demonstrate that the reconstruction process is successful whatever
the transversal position of the particle in the field of the beam. In addition, we show that the same fractional orders have to be
applied. We will thus be able to conclude that reconstruction is
possible for a whole wide-field object or wide particle field in
such a beam.

Now, to obtain the desired accuracy, we should analyze the number of
terms necessary in the series over $k$ in Eq. (\ref{fonction_A2})
and over $p$ in Eq. (\ref{fonction_Tk}). The series which should be
analyzed are in Eq.(\ref{Vnm}). The first upper bound that is
relevant here is \cite{Janssen04,Braat,Janssen02}
\begin{equation}\label{fonctionjm}
\left|j_{m}\small{\left(\frac{\gamma}{2}\right)}\right|\leq
\frac{1}{(2m+1)^{1/2}}\,\min
\left(1,\left(\frac{\pi}{2}\right)^{1/2}\frac{|\gamma/4|^{m}}{m!}\right).
\end{equation}
The coefficients $\omega_{ml}$ verify $\omega_{ml}\geq0$ and
$\sum_{l}\omega_{ml}=1$, so the $\omega$'s are completely innocent.
Using the bound (\ref{fonctionjm}) for the function
$\left|j_{m}\small{\left(\frac{\gamma}{2}\right)}\right|$ requires a
rough estimate of the variable $\gamma$. In practical experiments,
we have $D\approx 10^{-4}m$, $\lambda\approx 10^{-6}m$, $z\approx
10^{-1}m$, $R_{q}\approx 0.5\cdot 10^{-1}$ then $\gamma\approx
0.785\cdot 10^{-1}+0.816\cdot 10^{-3}i$. The right-hand side of Eq.
(\ref{fonctionjm}) is less than $0.142\cdot 10^{-1}$ for $m\geq1$.
The second upper bound we use concerns the Bessel functions
$J_{2k+2l+1}$. From \cite{Abramowitz}, 9.1.62 on p.362 and
\cite{Janssen04} we obtain:
\begin{equation}\label{fonctionbessel}
\left|\frac{J_{2k+2l+1}(r)}{r}\right|\leq \min\left(1,\frac{1}{2}
\frac{|r|^{2k+2l}\exp(|r|^{2})}{(2k+2l+1)!}\right).
\end{equation}
The evaluation of the accuracy concerns the product of the left-hand
side of Eqs. (\ref{fonctionjm}) and (\ref{fonctionbessel}):
\begin{equation}\label{precision}
\left|j_{m}\small{\left(\frac{\gamma}{2}\right)}\right|\left|\frac{J_{2k+2(m+p)+1}(r)}{r}\right|.
\end{equation}
With the previous values, all the quantities in (\ref{precision})
are less than $0.142\cdot 10^{-1}$ for all $(k,p,m)\geq 1$. Thus, we
only consider the case where $k=p=m=0$. In conclusion, the function
$V_{0}^{0}(r,\gamma)$ is given with good accuracy by:
\begin{equation}\label{fonctionVnm00}
V_{0}^{0}(r,\gamma)\simeq \exp(i\gamma/2)j_{0}\small{(\gamma/2)}
\frac{J_{1}(r)}{r}.
\end{equation}
Note that this result is true if the object is far from the beam
waist. The amplitude $A_2$ then becomes
\begin{equation}\label{holo:A2_simplifiee}
A_{2}=\frac{\pi D^2}{2\lambda z}\beta_{0}^{0}(\delta)\exp\left[\Phi
(\xi_{0},\eta_{0})\right]\exp\left(
ic_z\left[(x-\xi_0)^2+(y-\eta_0)^2\right]\right) \cdot
V_{0}^{0}(r,\gamma).
\end{equation}

\subsection{Intensity distribution of the diffraction pattern}

The intensity distribution of the diffraction pattern in the
quadratic sensor plane, denoted $I$, is evaluated from the Eqs.
(\ref{holo:equ4}), (\ref{holo:A2_simplifiee}) and (\ref{holo:equ5})
in the following way:
\begin{equation}\label{holo:equ16}
I=A\overline{A}=\left[A_{1}-A_{2}\right]\left[\overline{A_{1}}-\overline{A_{2}}\right]=
\left[|A_{1}|^{2}+|A_{2}|^{2}\right]-2\,\Re\left\{A_{1}\overline{A_{2}}\right\},
\end{equation}
where the overhead bar denotes the complex conjugate, $\Re$ denotes
the real part. Thus, the intensity distribution recorded by the CCD
sensor is described by the Eq. (\ref{holo:equ16}). As one sees from
the form of $I$, the first and second terms, \textit{i.e.},
$|A_{1}|^{2}$ and $|A_{2}|^{2}$, do not generate interference
fringes with a linear instantaneous frequency, denoted $f_{i}(x)$,
in the CCD plane, \textit{i.e.} \cite{Meck97}:
\begin{equation}
f_{i}(x)=\frac{1}{2\pi}\frac{\partial \varphi(x)}{\partial
x}=\frac{1}{2\pi}\frac{\partial \arg(|A_{1,2}|)}{\partial x}=0.
\end{equation}
But the third term exhibits a phase which is composed of a constant
and a linear instantaneous frequency. This fact is important because
the fractional Fourier transform is an effective operator for
analyzing a signal containing a linear instantaneous frequency
(linear chirp functions). From Eq. (\ref{holo:equ16}) we write
\begin{equation}
A_{1}\overline{A_{2}}=|A_{1}\overline{A_{2}}|\exp\left[i\arg\left(A_{1}\overline{A_{2}}\right)\right],
\end{equation}
where $\arg\left(A_{1}\overline{A_{2}}\right)=\phi -\phi_{0}$ with
\begin{equation}\label{A1A2:1}
\phi=c_z\left(x^{2}(M_{\xi}-1)+y^{2}(M_{\eta}-1)\right)+2c_z\left(x\xi_{0}+y\eta_{0}\right)-
\arg\left(\frac{J_{1}(r)}{r}\right),
\end{equation}
and
\begin{equation}\label{A1A2:2}
\phi_{0}=\frac{\Re(\gamma)}{2}+\Im(\Phi
(\xi_{0},\eta_{0}))+\arg\left(j_{0}\small{\left(\frac{\gamma}{2}\right)}\right)+\arg\left(\beta_{0}^{0}(\delta)\right)-\arg\left(K_\xi
K_\eta\right),
\end{equation}
where $\Im$ represent the imaginary part of a complex number. The
first term in (\ref{A1A2:1}) yields a quadratic phase and the second
term yields a linear phase. Remember that the aim of the
reconstruction by means of FRFT is precisely to analyze a linear
chirp.

To give two different examples, it is necessary to fix the values of
the parameters $(\omega_{\xi},\omega_{\eta})$, $(R_\xi,R_{\eta})$
and $(D,\lambda,z)$.

In the first case, the values are defined by $(7mm,1.75mm)$ for the
beam waists, $(-\infty,-50mm)$ for the wave's curvatures and the
diameter $D$ of the particle is equal to $150µm$ and located at
$120mm$ from the CCD sensor. The wavelength of the laser beam is
$632.8nm$. The distance between the cylindrical lens and the
particle is $\delta = 250mm$. The particle is shifted from the
origin by $(\xi_{0},\eta_{0})=(0.5mm,0.2mm)$. Figure (\ref{fig2})
illustrates the diffraction pattern which is recorded by the camera.
Note that the shift of the diffraction pattern $(x_0,y_0)$ observed
in the camera plane is not equal to the object shift
$(\xi_{0},\eta_{0})$. If the particle is considered far from the
beam waist then we have the formula:
\begin{equation}\label{thales}
y_0=\frac{\mid\Delta\mid\pm z}{\mid\Delta\mid}\,\,\eta_{0}.
\end{equation}
The sign of $z$ depends on the position of the particle compared with
the position of the waist. If the particle is after the waist, the
sign is positive. If it is in front of the waist, the sign is
negative. Along $x$-axis, the parameter $\Delta$ is infinite so that
in the plane of the camera $x_0=\xi_0=0.5mm$ and along $y$-axis,
$\Delta=50mm$ thus $y_0=0.68mm$.

Now, in the second case, the values are defined by $(7mm,1.75mm)$
for the beam waists, $(-\infty,50mm)$ for the wave's curvatures and
the diameter $D$ of the particle is equal to $150µm$ and located at
$z=120mm$ from the CCD sensor. The distance between the cylindrical
lens and the particle is $\delta = 150mm$. The particle is shifted
from the origin by $(\xi_{0},\eta_{0})=(0.5mm,0.2mm)$. Figure
(\ref{fig3}) illustrates the diffraction pattern which is recorded
by the camera. In the plane of the camera, $x_0=\xi_0=0.5mm$,
$\Delta=50mm$ thus, from Eq. (\ref{thales}), it leads to
$y_0=-0.28mm$. The diffraction pattern changes from elliptical
fringes to hyperbolic fringes. These diffraction patterns will be
used to reconstruct the image of the particle by FRFT.

\section{Fractional Fourier transformation analysis of in-line
holograms}

\subsection{ Two-dimensional Fractional Fourier transformation }

FRFT is an integral operator that has various application in
signal and image processing.
Its mathematical definition is
given in Ref. \cite{Namias80,McBride87,Lohmann93}. The
two-dimensional fractional Fourier transformation of order $a_{x}$
for $x$-cross-section and $a_{y}$ for $y$-cross-section with
$0\leq|\alpha_{x}|\leq\pi/2$ and $0\leq|\alpha_{y}|\leq\pi/2$,
respectively, of a 2D-function $I(x,y)$ is defined as (with
$\alpha_{p}=\frac{a_{p}\pi}{2}$)
\begin{equation}\label{FRFT:equ1}
\mathscr{F}_{\alpha_{x},\alpha_{y}}[I(x,y)](x_{a},y_{a})=\int_{\mathbb{R}^{2}}
N_{\alpha_{x}}(x,x_{a})\,N_{\alpha_{y}}(y,y_{a})I(x,y)\,dx\,dy,
\end{equation}
where the kernel of the fractional operator is defined by
\begin{equation}\label{FRFT:equ2}
N_{\alpha_{p}}(x,x_{a})=C(\alpha_{p})\exp\left(i\pi\frac{x^{2}+x_{a}^{2}}{s_p^{2}\tan
\alpha_{p}}\right) \exp\left(-\frac{i2\pi x_{a}
x}{s_p^{2}\sin\alpha_{p}}\right),
\end{equation}
and
\begin{equation}\label{FRFT:equ3}
C(\alpha_{p})=\frac{\exp(-i(\frac{\pi}{4}\textrm{sign}(\sin\alpha_{p})-\frac{\alpha_{p}}{2}))}
{|s_p^{2}\sin\alpha_{p}|^{1/2}}\,.
\end{equation}
Here $p=x,y$. Generally, the parameter $s_p$ is considered as a
normalization constant. It can take any value. In our case, its
value is defined from the experimental set-up according to
\cite{Mas}
\begin{equation}
s_p^2=N_p\cdot\delta_p^2.
\end{equation}
This definition is presented in the Appendix \ref{appC}. $N_p$ is
the number of samples along the $x$ and $y$ axes in both spatial
$I(x,y)$ and fractional domains. The constant $\delta_p$ is the
sampling period along the two previous axes of the image. In our
case the number of samples and the sampling period are the same
along both axes, so that the parameters $s_p$ are equal to $s$. The
energy-conservation law is ensured by the coefficient
$C(\alpha_{p})$ which is a function of the fractional order. One of
the most important FRFT feature is its ability to transform a linear
chirp into a Dirac impulse. Let $g(x)=\exp\left(i\pi\chi
x^2\right)$ the chirp function to be analyzed. If we consider $\frac{1}{\chi}=-\tan\alpha$, and using
the fact that:
\begin{equation}
\lim_{\varepsilon\rightarrow 0}\frac{1}{\sqrt{i\pi\varepsilon}}\exp\left(-\frac{x^2}{i\varepsilon}\right)=\delta(x),
\end{equation}
then the FRFT of optimal order $\alpha$ of $g(x)$  can be written
as:
\begin{equation}
\mathscr{F}_{\alpha}\left[g(x)\right]=\delta(x_a).
\label{dirac}
\end{equation}
Finally, by choosing the adequate value of the fractional order, a
pure linear chirp function can be transformed into a delta Dirac
distribution.

\subsection{Reconstruction: optimal fractional orders}

To reconstruct the image of the particle, the FRFT of the
diffraction pattern (Eq.(\ref{holo:equ16})) must be calculated:
\begin{align}\label{frftholo:equ4}
\begin{split}
\mathscr{F}_{\alpha_{x},\alpha_{y}}[I]&=\mathscr{F}_{\alpha_{x},\alpha_{y}}\left[|A_{1}|^{2}\right]
-\mathscr{F}_{\alpha_{x},\alpha_{y}}\left[2|A_{1}\overline{A_{2}}|
\cos(\phi-\phi_{0})\right]+\mathscr{F}_{\alpha_{x},\alpha_{y}}\left[|A_{2}|^{2}\right]
\end{split}
\end{align}
The terms $|A_{1}|^{2}$ and $|A_{2}|^{2}$ do not contain a linear
chirp, so they do not have any effect on the optimal fractional
order to be determined. But the second term, denoted $S_{t}$ contains a
linear chirp. It will be considered for the image reconstruction of
the particle. By noting that
$2\cos(\phi-\phi_{0})=\exp(-i(\phi-\phi_{0}))+\exp(i(\phi-\phi_{0}))$,
the second term of Eq. (\ref{frftholo:equ4}) becomes :
\begin{equation}
\mathscr{F}_{\alpha_{x},\alpha_{y}}\left[2|A_{1}\overline{A_{2}}|
\cos(\phi-\phi_{0})\right]=\exp\left(i\pi\frac{x_{a}^{2}}{s^{2}\tan\alpha_{x}}\right)\exp\left(i\pi\frac{y_{a}^{2}}{s^{2}\tan\alpha_{y}}\right)
\left\{I_{-}+I_{+}\right\}\label{idec}
\end{equation}
with
\begin{equation}
I_{\pm}=C(\alpha_x)C(\alpha_y)\iint_{\mathbb{R}^2}\left|A_1\overline{A_2}\right|\exp\left[i\left(\phi_a\pm(\phi-\phi_{0})\right)\right]\exp\left[-\frac{2i\pi}{s^2}\left(\frac{x_ax}{\sin\alpha_x}+\frac{y_ay}{\sin\alpha_y}\right)\right]dxdy
\end{equation}
The quadratic phase term of the FRFT is denoted by $
\phi_a=\frac{\pi}{s^2}\left(x^2\cot\alpha_x+y^2\cot\alpha_y\right)$.
Let us recall that the FRFT allows us to analyze a linear chirp.
Thus, if the fractional orders check the following conditions
\begin{equation}\label{optimalorders}
\frac{\pi\cot\alpha_{x}^{opt}}{s^{2}}=c_z(M_x-1),\quad \quad
\frac{\pi\cot\alpha_{y}^{opt}}{s^{2}}=c_z(M_y-1),
\end{equation}
then, the FRFT of $I_{-}$ is no more than a classical Fourier
transformation according to
\begin{equation}
I_-=\chi\cdot\mathcal{F}\left[\frac{J_{1}(r)}{r}\cdot\exp\left(-\frac{\pi}{\lambda
z}\rho^{\mathrm{T}}N \rho\right)\right](u,v),
\end{equation}
with $\chi=\frac{\pi D^2}{2\lambda z}C(\alpha_x)C(\alpha_y)K_\xi
K_\eta \exp\left[\Phi
(\xi_{0},\eta_{0})\right]\beta_{0}^{0}(\delta)\exp\left(i\frac{\gamma}{2}\right)j_{0}\small{\left(\frac{\gamma}{2}\right)}$.
The operator $\mathcal{F}$ is the 2D-Fourier transformation. The
spatial frequencies $u$ and $v$ are equal to:
\begin{equation}
u=\frac{x_{a}}{s^{2}\sin(\alpha_{x}^{opt})}+\frac{c_z\xi_0}{\pi}
\quad \quad
v=\frac{y_{a}}{s^{2}\sin(\alpha_{y}^{opt})}+\frac{c_z\eta_0}{\pi}
\end{equation}
As the Fourier transform of the product of two functions is equal to
the convolution of their transforms then
\begin{equation}\label{convolution}
I_-=\chi\cdot\mathcal{F}\left[\frac{J_{1}(r)}{r}\right]\ast
\mathcal{F}\left[\exp\left(-\frac{\pi}{\lambda z}\rho^{\mathrm{T}}N
\rho\right)\right],
\end{equation}
Remember here that the variables $\rho$ and $r$ pertain to the same
coordinates $(x,y)$. With the shift theorem for the Fourier
transform, the Hankel transform and the discontinuous
Weber-Schafheitlin integral in [\cite{Abramowitz}], 11.4.42 on
p.487, we have:
\begin{equation}\label{fonction_circ}
\mathcal{F}\left[\frac{J_{1}(r)}{r}\right]=2\pi \left(\frac{\lambda
z}{\pi D}\right)^{2}\exp\left[-i2\pi(uX_{0}+vY_0)\right]\times
  \begin{cases}
    1, & \text{$0<\sqrt{u^{2}+v^{2}}$ $<$ $\frac{D/2}{\lambda z}$}, \\
    1/2, & \text{$0<\sqrt{u^{2}+v^{2}}$ = $\frac{D/2}{\lambda z}$}, \\
    0, & \text{$\sqrt{u^{2}+v^{2}}$ > $\frac{D/2}{\lambda z}>0$},
  \end{cases}
\end{equation}
with $X_0=\xi_{0}\left(1-ic_\xi/c_z\right)$ and
$Y_0=\eta_{0}\left(1-ic_\eta/c_z\right)$. The function defined by
the right-hand side of Eq. (\ref{fonction_circ}), with spatial
coordinates $(x_a,y_a)$, has the aperture of the pinhole with
diameter equal to the diameter $D$ of the opaque particle. We have
thus demonstrated our result: reconstruction with a FRFT leads
exactly to the object. In addition, shifting the object does
not modify the fractional order. This point is important because in
the case of a particle field, a single fractional order couple
along the $x-$ axis and $y-$ axis is necessary to reconstruct the
particle image. If one wishes to determine the shift of the
diffraction patterns in the $(x_a,y_a)$-plane, the coordinates that
should be considered are:
\begin{equation}
\left(s^2(u-\frac{c_z\xi_{0}}{\pi})\tan\alpha_{x}^{opt},
s^2(v-\frac{c_z\eta_{0}}{\pi})\tan\alpha_{y}^{opt}\right)
\end{equation}
This correction is necessary because the fractional Fourier
transformation is not invariant by translation: a space-shift in the
spatial domain will lead to both frequency-shift and phase-shift in
the FRFT domain. The shift rule of the FRFT is expressed as:
\begin{equation}
\mathscr{F}_{\alpha}[f(x-b)](u)=\exp\left(i\pi
b^2\sin\alpha\cos\alpha\right)\exp\left(-i2\pi
bu\sin\alpha\right)\mathscr{F}_{\alpha}[f(x)](u-b\cos\alpha)
\end{equation}
Note that the nature of the Gaussian beam implies that the
reconstructed image of the object is convolved by a 2D Gaussian
function. The sign of the object function (opaque object with a
luminous background) is retrieved by applying an inversion of
$I_{-}$ that is realized by the minus in front of the second term of
Eq. (\ref{frftholo:equ4}).

\subsection{Numerical experiments}
\begin{figure}[t]
\centering
\includegraphics*[width=15cm]{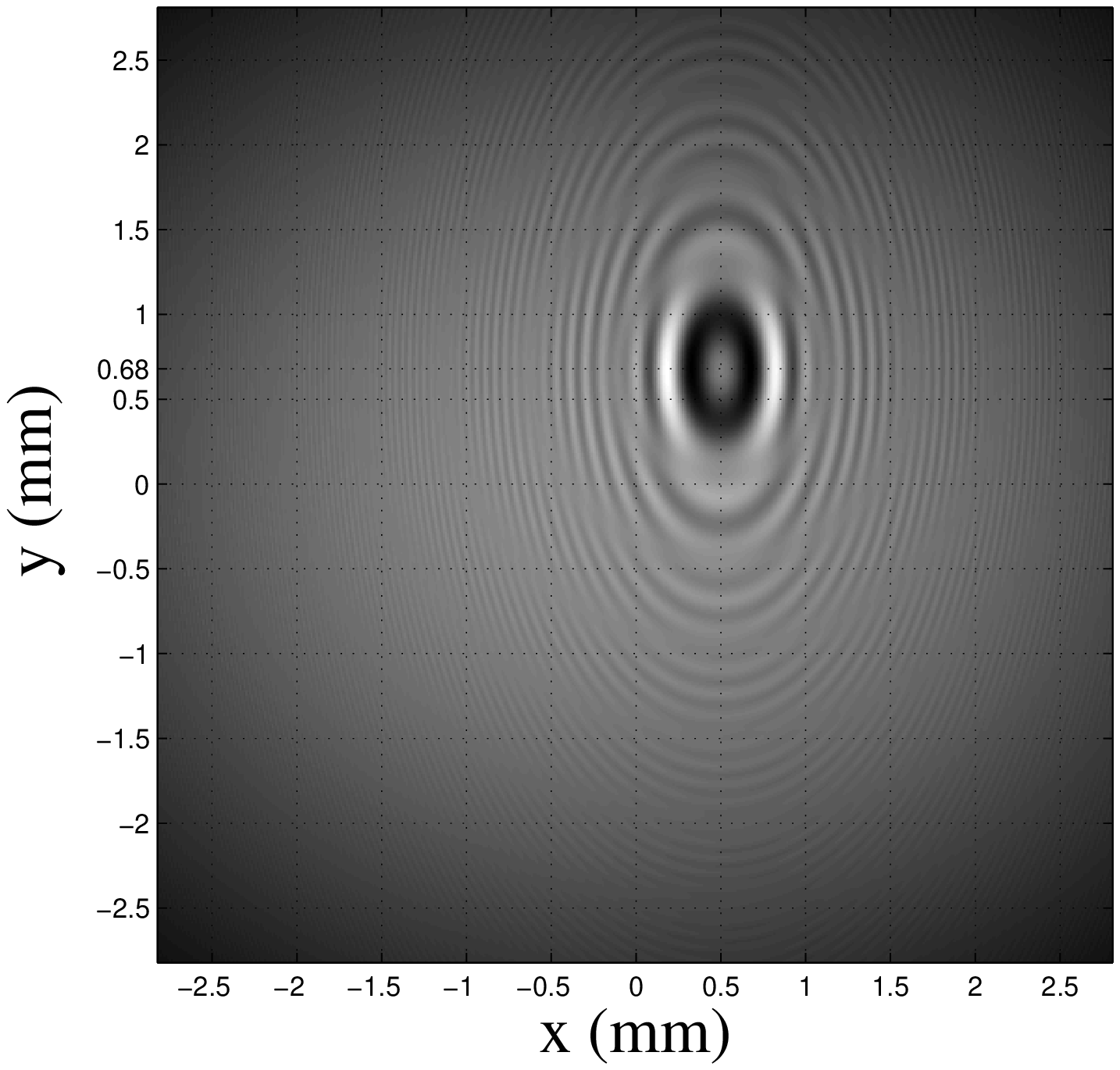}
\caption{Diffraction pattern with , $\omega_{\xi}=7mm$,
$\omega_{\eta}=1.75mm$, $R_{\xi}=\infty$, $R_{\eta}=-50mm$,
$D=150µm$, $\lambda=632.8nm$, $z=120mm$, $\delta=250mm$,
$\xi_{0}=0.5mm$ and $\eta_{0}=0.2mm$.}\label{fig2}
\end{figure}

\begin{figure}[t]
\centering
\includegraphics*[width=15cm]{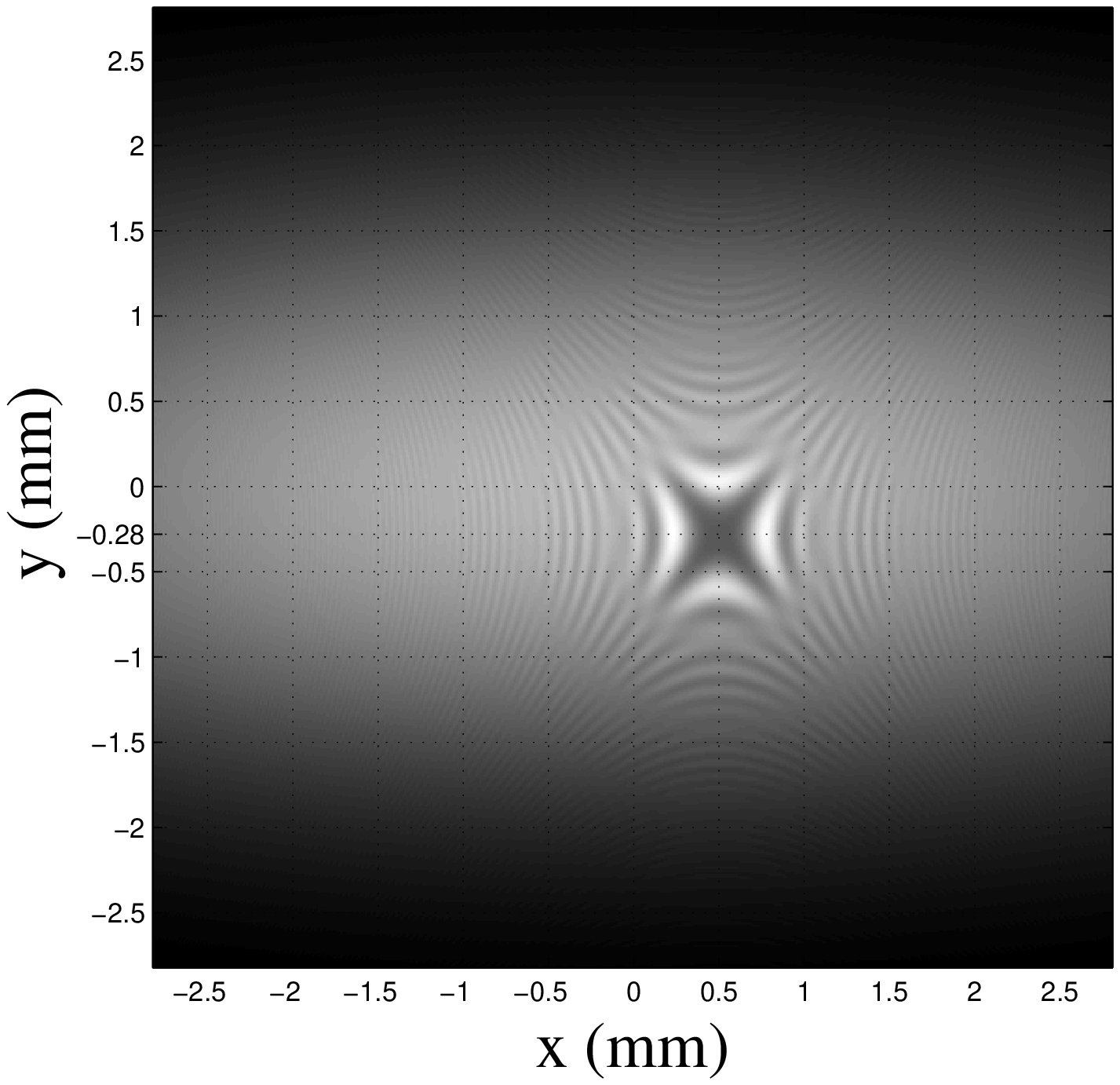}
\caption{Diffraction pattern with , $\omega_{\xi}=7mm$,
$\omega_{\eta}=1.75mm$, $R_{\xi}=\infty$, $R_{\eta}=-50mm$,
$D=150µm$, $\lambda=632.8nm$, $z=120mm$, $\delta=150mm$,
$\xi_{0}=0.5mm$ and $\eta_{0}=0.2mm$.}\label{fig3}
\end{figure}

\begin{figure}[t]
\centering
\includegraphics*[width=15cm]{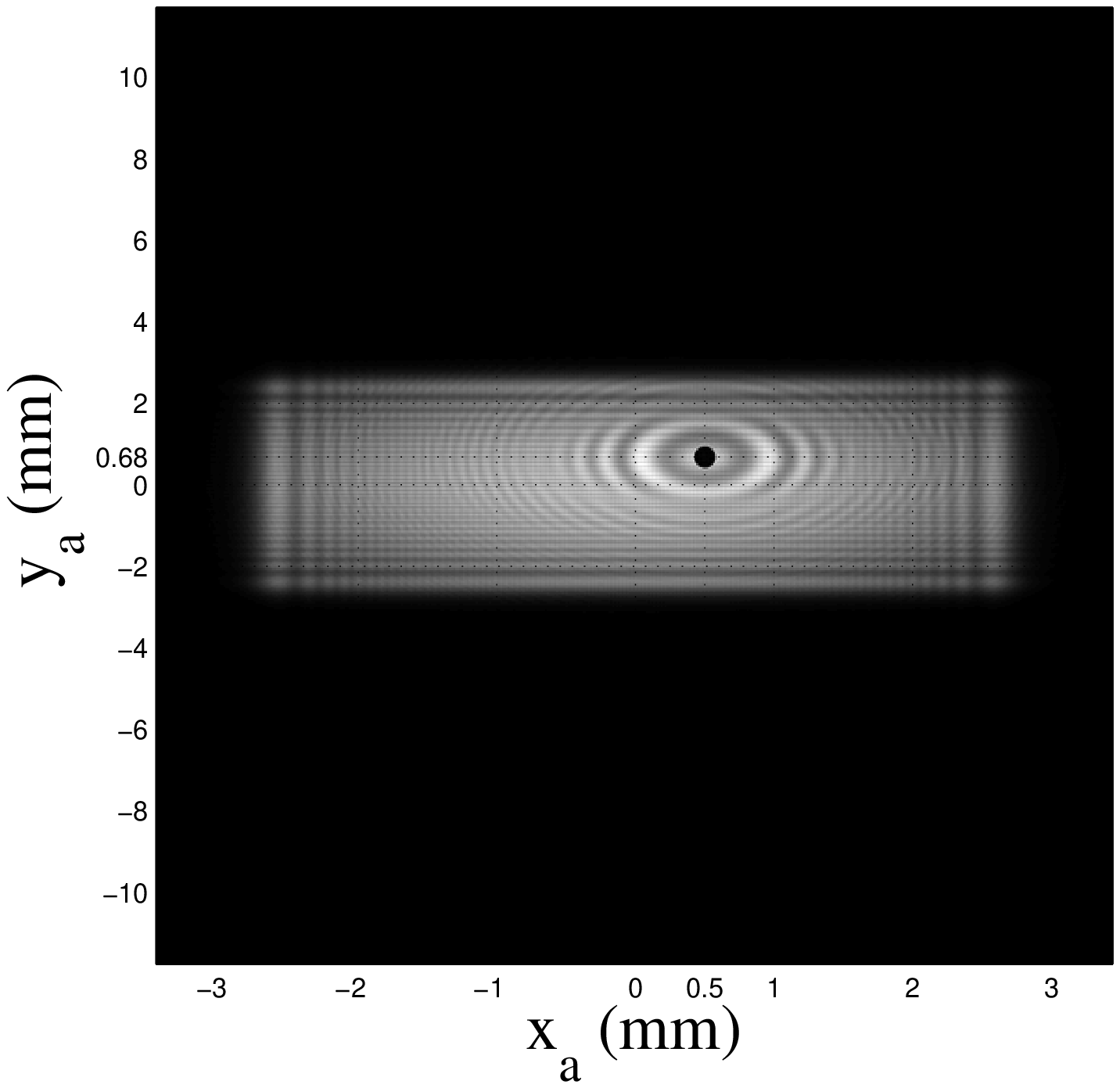}
\caption{Fractional Fourier transform of the diffraction pattern
with $a_x^{opt}=-0.564$ and $a_y^{opt}=-0.850$.}\label{fig4}
\end{figure}

\begin{figure}[t]
\centering
\includegraphics*[width=15cm]{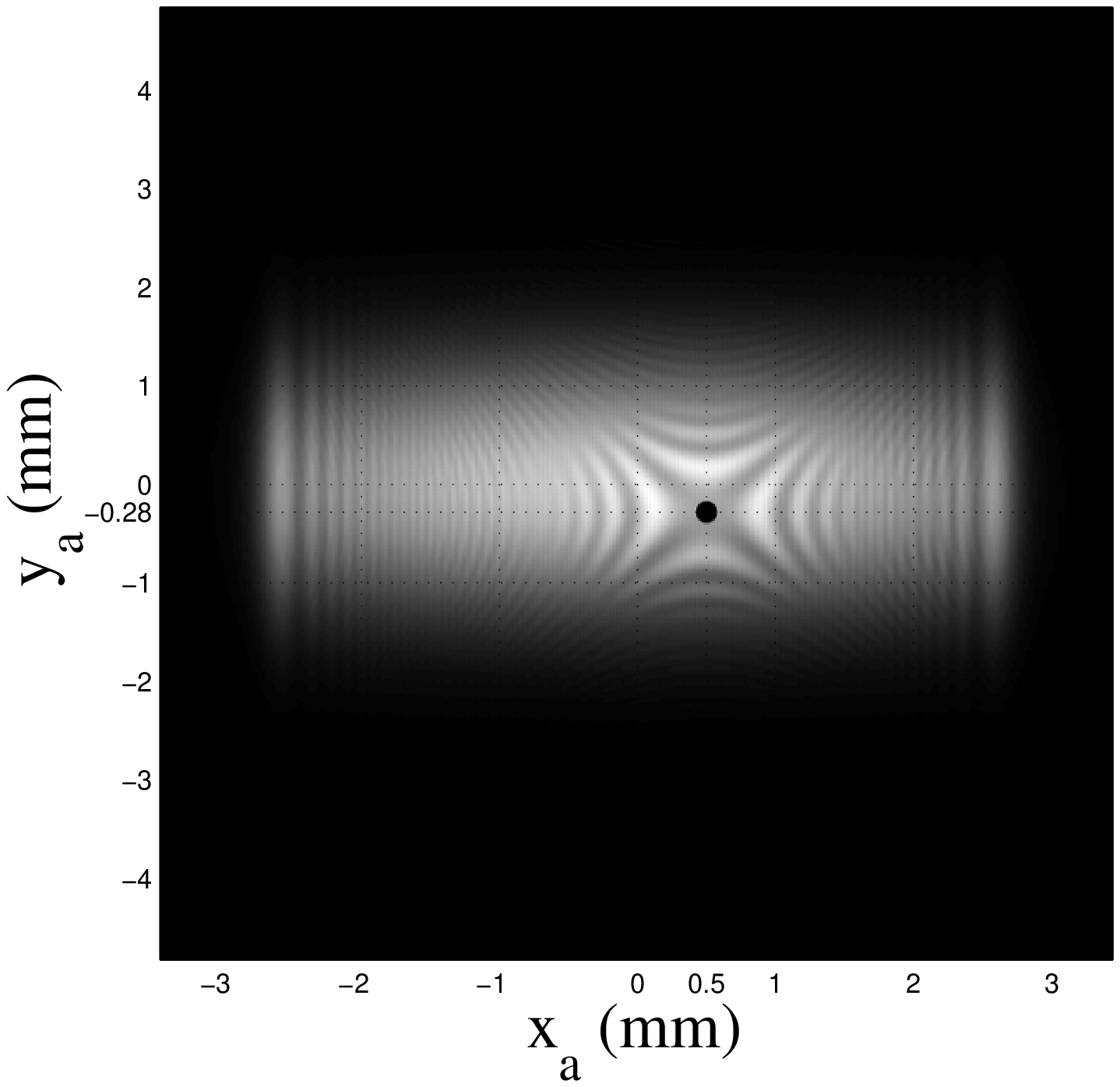}
\caption{Fractional Fourier transform of the diffraction pattern
with $a_x^{opt}=-0.564$ and $a_y^{opt}=0.664$.}\label{fig5}
\end{figure}

The simulations of the particle image reconstruction are realized
from the diffraction patterns illustrated by the Figs. (\ref{fig2})
and (\ref{fig3}). The diffraction patterns consist of a $512\times
512$ array of $11µm\times 11µm$ size pixels. Consider the
diffraction pattern presented in Fig. (\ref{fig2}) and produced by a particle of $D=150µm$ in diameter located at $z=120mm$. The optimal fractional
orders obtained from Eq. (\ref{optimalorders}) are
$a_x^{opt}=-0.564$ and $a_y^{opt}=-0.850$. The image of the
reconstructed image is shown in Fig. (\ref{fig4}). In this
representation, the squared modulus of the FRFT, \textit{i.e.}
$|\mathscr{F}_{\alpha_{x},\alpha_{y}}[I]|^{2}$, is taken. The shape
of the particle image is not modified: the width of the Gaussian
function in the Eq.(\ref{convolution}) is greater than the diameter
$D$ of the particle (typically $679µm$ along x-cross axis and
$490µm$ along y-cross axis). Now, the reconstruction of the particle
image from the diffraction pattern illustrated by Fig. (\ref{fig3}),
is realized by a fractional Fourier transformation of optimal orders
$a_x^{opt}=-0.564$ and $a_y^{opt}=0.664$. Figure (\ref{fig5})
illustrates the result of the reconstruction. In both cases
(Fig.(\ref{fig4}) and Fig.(\ref{fig5})) reconstruction by FRFT is
successful. Note that we have checked that the apertures resulting
from Eq. (\ref{convolution}) and the image of the reconstructed
particle give the same diameter $D$.

\subsection{Experimental results}

As the reconstruction process is successful whatever the transversal
position of the particle in the field of the beam, for the same
fractional orders we can now carry out a reconstruction for a
wide-field object.

\begin{figure}[t]
\centering
\includegraphics*[width=15cm]{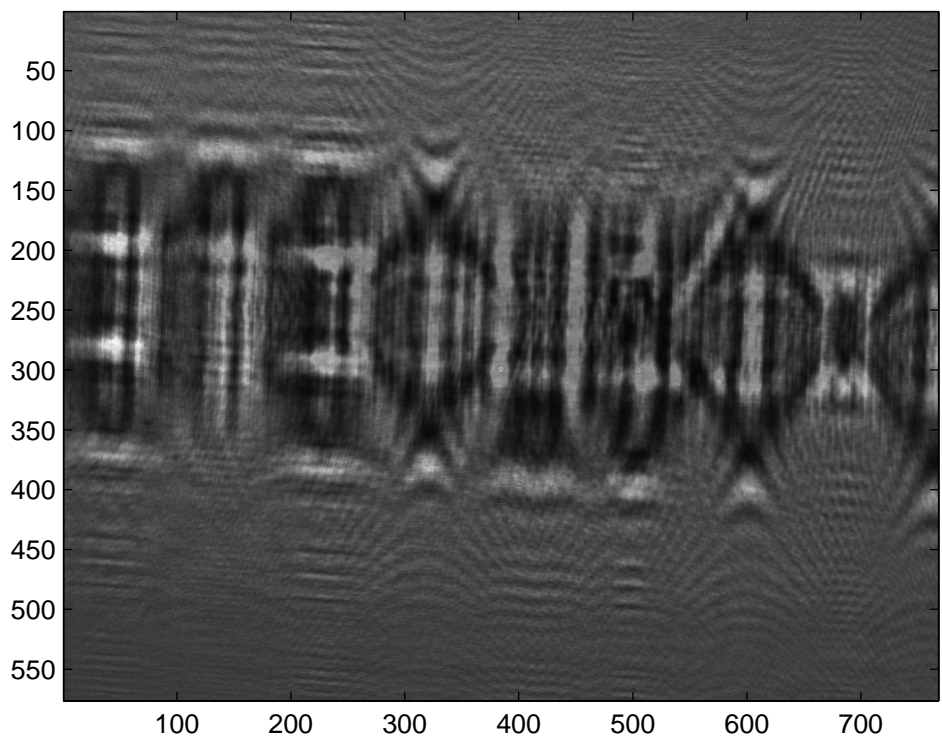}
\caption{Diffraction pattern of the word "ELECTRO".}\label{fig6}
\end{figure}

\begin{figure}[t]
\centering
\includegraphics*[width=15cm]{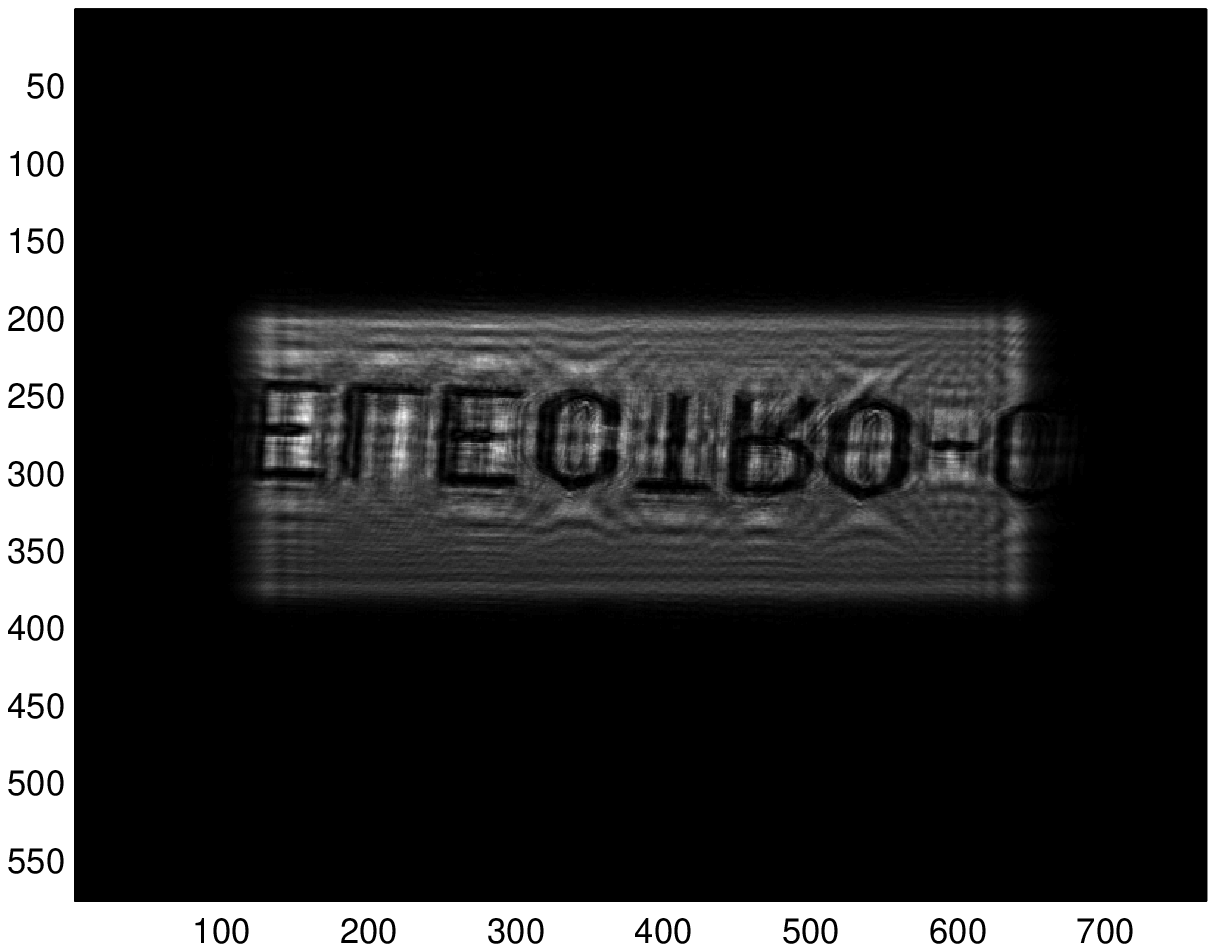}
\caption{Fractional Fourier transform of the diffraction pattern
with $a_x^{opt}=0.505$ and $a_y^{opt}=-0.785$.}\label{fig7}
\end{figure}

The previous theoretical developments and numerical experiments have
been tested by using an RS-3 standard reticle (Malvern Equipment).
This reticle is an optical glass plate with a pattern of the word
"ELECTRO" photographically deposited on the surface. The word
"ELECTRO" spread over $6.5mm$. The reticle is located at
$\delta\approx156mm$ from the cylindrical lens (CL). The distance
$z$ between the CCD camera and the reticle is approximatively equal
to $117mm$. The opaque word "ELECTRO" is in front of the waist of
the beam. The intensity distribution of the EAGB diffracted by the
word "ELECTRO" is shown in Fig. (\ref{fig6}). The image of the word
"ELECTRO" is reconstructed by the fractional Fourier transformation.
Fractional orders have been adjusted to obtain the best contrast
between the reconstructed "ELECTRO" word and the background. Thus
doing, the approximate orders are $a_x^{opt}=0.505$ and
$a_y^{opt}=-0.785$. The image of Fig. (\ref{fig7}) shows that two
previous orders allow all parts of the image of the object to be
reconstructed. Note that the word is reversed along the $y-$axis.
This is due to the Gouy phase shift of a Gaussian beam that
propagates from $-\infty$ to $+\infty$ through the focus point. In
this case, the phase shift predicted is $\pi$.

\section{Conclusion}
The effect of a transversal object shift  in an elliptical,
astigmatic Gaussian beam does not affect the optimal fractional
orders required to reconstruct the image of a particle or any other
opaque object in the field. In this publication an analytical model
has been developed to prove this. In this development, an opaque
particle is considered. Experimental result presented by the
recording and reconstruction of the word "ELECTRO" show that our
method works, even when the object is spread out over the whole
image field.

\newpage
\appendix

\section{Appendix A : Expression of the terms $A_1$ and $A_2$}\label{appA}

\subsection{Expression for the amplitude distribution  $A_{1}$ }

The development of integral $A_{1}$ has been given in a previous
paper by \cite{Nicolas} :
\begin{equation}\label{holo:equ5}
A_{1}=K_\xi K_\eta\exp\left(-\frac{\pi}{\lambda z}\rho^{\mathrm{T}}N
\rho\right) \exp\left(i\frac{\pi}{\lambda z} \rho^{\mathrm{T}} M
\rho\right)
\end{equation}
where $\rho^{\mathrm{T}}$ represents the vector $(x \quad y)$ and
the factors $K_q$ with $q=\xi, \eta$ in Eq.(\ref{holo:equ5}) are
defined by
\begin{equation}\label{holo:equ6}
K_q=\left[\frac{\frac{\pi \omega_{q}^{2}}{\lambda z}}{1+i\frac{\pi
\omega_{q}^{2}}{\lambda
z}\left(\frac{z}{R_{q}}-1\right)}\right]^{1/2}
\end{equation}
and the diagonal matrices $N$ and $M$ by
\begin{equation}\label{holo:equ7}
N=\left(%
\begin{array}{cc}
  N_{x} & 0 \\
  0 & N_{y} \\
\end{array}%
\right), \quad\quad M=\left(%
\begin{array}{cc}
  M_{x} & 0 \\
  0 & M_{y} \\
\end{array}%
\right),
\end{equation}
with
\begin{equation}\label{holo:equ8}
N_{q}=\pi\,\frac{\frac{\omega_{q}^{2}}{\lambda
z}}{1+\pi^{2}\frac{\omega_{q}^{4}}{(\lambda
z)^{2}}\left(\frac{z}{R_{q}}-1\right)^{2}} \quad\quad
\textrm{,}\quad\quad
M_{q}=1+\pi^{2}\,\,\frac{\,\frac{\omega_{q}^{4}}{(\lambda
z)^{2}}\left(\frac{z}{R_{q}}-1\right)}
 {1+\pi^{2}\frac{\omega_{q}^{4}}{(\lambda
z)^{2}}\left(\frac{z}{R_{q}}-1\right)^{2}}.
\end{equation}

\subsection{Expression for the amplitude distribution  $A_{2}$ }

To develop the second integral of $A_2$, involving the product of
$E(\xi,\eta)$ and $T(\xi-\xi_{0},\eta-\eta_{0})$, \textit{i.e.}:
\begin{equation}
A_{2}=\frac{\exp\left[\frac{i\pi}{\lambda
z}(x^{2}+y^{2})\right]}{\lambda z}\int_{\mathbb{R}^2}
E(\xi,\eta)T(\xi-\xi_{0},\eta-\eta_{0}) \exp\left[\frac{i\pi
}{\lambda z}(\xi^{2}+\eta^{2})\right]\exp\left[-i\frac{2\pi}{\lambda
z}(x\xi+y\eta)\right]\,d\xi d\eta,
\end{equation}
we first replace $\xi$ by $\xi+\xi_0$ and $\eta$ by $\eta+\eta_{0}$
and get:
\begin{multline}
A_{2}=\frac{\exp\left[\frac{i\pi}{\lambda
z}(x^{2}+y^{2})\right]}{\lambda z}\int_{\mathcal{D}}
E(\xi+\xi_{0},\eta+\eta_{0}) \exp\left[\frac{i\pi }{\lambda
z}((\xi+\xi_{0})^{2}+(\eta+\eta_{0})^{2})\right]\times\\\exp\left[-i\frac{2\pi}{\lambda
z}(x(\xi+\xi_{0})+y(\eta+\eta_{0}))\right] \,d\xi d\eta.
\end{multline}
The domain $\mathcal{D}$ must be defined as the disk with center $0$
and diameter $D$. By considering that $c_{z}=\pi/(\lambda z)$ and by
restating $A_{2}$ in cylindrical coordinates as follows:
$\xi=D\sigma\cos(\varphi)/2 $ and $\eta=D\sigma\sin(\varphi)/2$ for
the object plane, we obtain:
\begin{multline}
A_{2}=\frac{D^2}{4\lambda
z}\exp\left[c_{\xi}\xi_{0}^{2}+c_{\eta}\eta_{0}^{2}+
ic_z\left[(x-\xi_0)^2+(y-\eta_0)^2\right]\right]\cdot\\
\int_{0}^{1}\int_{0}^{2\pi}\exp\left[i\gamma
\sigma^{2}\right]\exp\left[i\delta\sigma^{2}\cos(2\varphi)\right]
\exp\left[ia\sigma\cos\varphi+ib\sigma\sin\varphi\right]\,\sigma
d\sigma d\varphi
\end{multline}
with
\begin{eqnarray}
    \begin{array}{l}
      \displaystyle
      \gamma=\frac{D^2}{4}\,c_{z}-i\frac{D^2}{8}\,(c_{\xi}+c_{\eta}), \quad\quad \delta=i\,\frac{D^{2}}{8}\,(c_{\eta}-c_{\xi}), \\[0.4cm]
      \displaystyle
      a=Dc_z\left[\xi_{0}\left(1-ic_\xi/c_z\right)-x\right], \quad\quad b=Dc_z\left[\eta_{0}\left(1-ic_\eta/c_z\right)-y\right].\\[0.4cm]
      \end{array}
\end{eqnarray}
By writing:
\begin{equation}
a\cos\varphi+b\sin\varphi=r\cos(\varphi-\theta)
\end{equation}
for which we have the condition
\begin{equation}\label{condition}
a=r\cos\theta, \quad\quad b=r\sin\theta
\end{equation}
with complex $r$ and $\theta$. This representation is discussed in
some detail in Appendix \ref{appB}. By means of the following
equalities in [\cite{Abramowitz}], 9.1.41 - 9.1.45:
\begin{equation}
\exp\left[i\delta\sigma^{2}\cos\left(2\varphi+2\theta\right)\right]=J_{0}\left(\delta\sigma^{2}\right)+
2\sum_{k=1}^{+\infty}i^{k}
J_{k}\left(\delta\sigma^{2}\right)\cos2k(\varphi+\theta),
\end{equation}
and
\begin{equation}
\frac{1}{2\pi}\int_{0}^{2\pi}\exp(in\theta)\exp[ix\cos\theta]d\theta=i^{n}J_{n}(x),
\end{equation}
the expression of $A_2$ becomes:
\begin{equation}\label{holo:A2}
A_{2}=\frac{\pi D^2}{\lambda z}\exp\left[\Phi
(\xi_{0},\eta_{0})\right]\cdot \exp\left(
ic_z\left[(x-\xi_0)^2+(y-\eta_0)^2\right]\right) \cdot
\sum_{k=0}^{\infty}(-i)^{k}\varepsilon_{k}\,T_{k}(r,\gamma)\cos(2k\theta),
\end{equation}
with $\varepsilon_{k}=1/2$ if $k=0$ and $1$ otherwise. The parameter
denoted $\Phi (\xi_{0},\eta_{0})$ is equal to
$\left[c_{\xi}\xi_{0}^{2}+c_{\eta}\eta_{0}^{2}\right]$. The function
$T_{k}(r,2\gamma)$ is defined as:
\begin{equation}\label{holo:equ14}
T_{k}(r,\gamma)=\sum_{p=0}^{\infty}\beta_{2k+2p}^{2k}(\delta)V_{2k+2p}^{2k}(r,\gamma),
\end{equation}
where the coefficients $\beta_{2k+2p}^{2k}$ are given by the
analytical development of $T_{k}$ in Appendix of [\cite{Nicolas}].
Recall here that the expression of $V_{2k+2p}^{2k}(r,\gamma)$ is:
\begin{equation}\label{fonctionVnm}
V_{2k+2p}^{2k}(r,\gamma)=\exp(i\gamma/2)\sum_{m=0}^{\infty}(2m+1)i^{m}
j_{m}\small{(\gamma/2)}\cdot
\sum_{l=\max(0,m-2k-p,p-m)}^{m+p}(-1)^{l}\omega_{ml}\frac{J_{2k+2l+1}(r)}{r}.
\end{equation}
The coefficients $\omega_{ml}$ is given explicitly in
[\cite{Nicolas}] and [\cite{Janssen04}].

\section{Appendix B : Elaboration of condition (\ref{condition})}\label{appB}
With $u=a+ib$, $v=a-ib$, we should find $\tau$ and
$\omega=\exp(i\theta)$ such that
\begin{equation}\label{equA1}
\tau \omega=u, \quad\quad \tau/\omega=v
\end{equation}
Assume that $u\neq 0$, $v \neq 0$, and write $u=r\exp(i\alpha)$,
$v=s\exp(i\beta)$ with $r,s>0$ and $\alpha, \beta\in \mathbb{R}$
then we see that $\omega=\exp(i\theta)$ given by
\begin{equation}
\tau=(rs)^{1/2}\exp\left(i(\alpha+\beta)/2\right), \quad\quad
\omega=\exp(i\theta)=\left(\frac{r}{s}\right)^{1/2}\exp\left(i(\alpha-\beta)/2\right)
\end{equation}
satisfy the relations (\ref{equA1}). This does not work in the case
that  $u=0$ or $v=0$. Indeed, when $a=1$, $b=i$ we get from
(\ref{condition}) and $\cos^{2}\theta+\sin^{2}\theta=1$ that
$r=a+ib=0$, \textit{i.e.}, $r=0$.

\section{Appendix C : Definition of $s_p$}\label{appC}

To determine the value of $s_p$, it is necessary to write the
definition of the one-dimensional fractional Fourier transformation
in the particular case of $\alpha=\pi/2$:
\begin{equation}\label{eqB1}
\mathcal{F}_{\pi/2}[I(x)](x_{a})= C(\pi/2)
 \int_{-\infty}^{+\infty}I(x)
\exp\left(-i2\pi\frac{x x_{a}}{s^{2}}\right)\,dx.
\end{equation}
Its discrete version is
\begin{equation}\label{eqB2}
\mathcal{F}_{\pi/2}[I(m)](k)=C(\pi/2)
\sum_{m=-N/2}^{N/2-1}I(m)\exp\left(-i2\pi\frac{m\delta _{x} \ k
\delta _{x_{a}} }{s^{2}}\right)\delta_{x},
\end{equation}
where $\delta_{x}$ and $\delta_{x_{a}}$ are the sampling periods of
$I(x)$ and its transform. The sampling periods are equal to $\delta
_{x}$, and $N$ is the number of samples of $I(x)$ and its transform.
The relation (\ref{eqB2}) can be written as the discrete Fourier
transformation of $I(m)$ according to:
\begin{equation}\label{eqB3}
\mathcal{F}_{\pi/2}[I(m)](k)=C(\pi/2)
\sum_{m=-N/2}^{N/2-1}I(m)\exp\left(-i2\pi\frac{mk }{N}\right)\delta
_{x}.
\end{equation}
By identification of Eqs. (\ref{eqB3}) and (\ref{eqB2}), one
obtains:
\begin{equation}\label{eqB4}
\frac{\delta _{x} \delta _{x_{a}}}{s^{2}}=\frac{1}{N}\quad\quad
\textrm{so}\quad\quad s^{2}=N \delta _{x} \delta
_{x_{a}}=N\delta_{x}^{2}.
\end{equation}
In the case of two-dimensional function one finally has
$s_p^2=N_p\cdot\delta_p^2$ with $p=\xi,\eta$.


\newpage
\listoffigures
\newpage

\end{document}